\documentclass{ifacconf}
\usepackage{enumerate}
\usepackage{wrapfig}
\usepackage{amsmath}
\usepackage{mathrsfs}
\usepackage{url}
\usepackage{booktabs}
\usepackage[T1]{fontenc}  
\usepackage{array, diagbox}

\usepackage{amsfonts,amsmath,amssymb}
\usepackage{graphicx}      
\usepackage{natbib}        
\begin{document}

\begin{frontmatter}

\title{State Estimation for Vision-based Localization under Uncertain Conditions} 
%
\thanks[footnoteinfo]{Both the authors contributed equally for this submission}

\author[First]{Prashant V. Patil \thanksref{footnoteinfo}}  
\author[Second]{Pranav Thakkar \thanksref{footnoteinfo}} 
\author[Third]{Leena Vachhani}
\address[First]{IITB-Monash Research Academy, Mumbai, India (e-mail: prashant\_patil@sc.iitb.ac.in)}
\address[Second]{Dept. of Aerospace Engineering, IIT Bombay, Mumbai, India (e-mail: pthakkar.2197@gmail.com)}
\address[Third]{Systems and Control Engineering, IIT Bombay, Mumbai, India (e-mail: leena.vachhani@iitb.ac.in)}

\begin{abstract}                
  Vision based localization is a popular approach to carry out manoeuvres  particularly in GPS-restricted indoor environments, because vision can complement other activities performed by the robot. The objective is to estimate the current location with respect to a known location by matching the bearings. The problem is challenging as the known location information is in terms of the bearings of landmarks extracted from an image. We address the problem under more challenging scenario when landmarks are semi-static. 
  In this work, an observer formulation is presented which enables to incorporate the effect of change in landmark position as parameters. The efficacy of two estimators: Augmented Extended Kalman Filter (A-EKF) and a Proportional-Integral EKF (PI-EKF) is tested under the cases where there are changes in some of the landmark positions. Morever, it is likely that not all landmarks are visible to the robot at all instants of time. A multi-rate estimation framework is proposed to mitigate this issue. Observability analysis is carried out to arrive upon a minimum number of landmarks required for such a formulation. Simulation studies are presented to test the efficacy of the formulations. 

\end{abstract}


\end{frontmatter}

\section{Introduction}
Navigation using bearing only information from the landmarks, is a  popular research area, with applications ranging from manoeuvres to docking as well as pick-and-place tasks. 
Vision based localization and navigation is a prominent approach in GPS restricted indoor environments (\cite{liu2010bearing}, \cite{corke2003mobile},\cite{bekris2006exploiting}). Vision based sensors can complement in dealing with the other activities performed by the robot, and this is what makes these sensors appealing for low cost applications.
Bearing information can be easily and reliably extracted from an image and a significant research progress has been achieved in this area. For the control formulation used in this exercise, it can be shown that this choice of measurements also eliminates the need to measure the position of the robot with respect to the home position instantaneously (\cite{ArunkumarG.K.2018}). However, an environment like a warehouse, or in a pick and place application, the landmarks could be typically be the products and items in the environment. It may happen that these landmarks are slightly disturbed from their position. Most of the popular localization methods consider the features of the landmarks as parameters. Thus, a change in the position of landmarks change could propagate a parametric change in the models. \cite{Rosen_Semi_Static_Environmet}, \cite{Delius_Semi_Static_Environment2} have proposed ways to mitigate the problems that arise in a semi-static environment. However, in our case, as the landmarks could be slightly disturbed from their positions, the approaches could be an overkill. 
Frameworks which may increase the computational burden on a low cost platform are also not desirable. Also, knowingly or otherwise, uncertainty seeps in the localization problem for a wheeled mobile robot, from sources other than those resulting from the disturbed landmarks. 
\\ The Bayesian state
estimation approach handles the effect of random noise in the states and the output of the system systematically, and hence is a widely used framework in this area(\cite{Prob_Robotics}).
The popular Bayesian estimation algorithms work satisfactorily provided the model parameters values are fairly accurately known. (\cite{PATWARDHAN2012933}). The accuracy of these Bayesian algorithms are leveraged on the accuracy of the model that is used to carry out the predictions. Using an inaccurate model or a model without correct parameter values could result in biased state estimates (\cite{BAVDEKAR2013184}). This can subsequently have ramifications on applications which use these estimates directly such as in state-feedback control formulations. It therefore becomes necessary to systematically take the parametric model-plant mismatch into account and modify the standard estimation algorithms. \\ 
Some of the widely popular approaches to compensate for this mismatch are the joint state and parameter estimation approaches, dual estimation approaches, adaptive estimation based approaches. The dual and the joint state-parameter estimation based approaches deal with the errors arsing from this mismatch by re-identifying the model parameters. The Proportional-Integral Kalman Filter(PI-KF) uses the idea of compensating the error caused due to the model-plant mismatch or by linearization, by a bias term, which is integrated over time based on the innovation sequence (\cite{BAVDEKAR2013184}). 
We propose a simple extension to this filter to suit the non-linear state transition map, and call it the Proportional-Integral Extended Kalman Filter(PI-EKF). The PI-EKF implementation is similar to the conventional Extended Kalman Filter(EKF). 
This nonlinear filter appears to be a promising, computationally efficient way of dealing with uncertainties and we use this filter in our studies and benchmark its results against the conventional EKF, and the Augmented State-Parameter estimation based EKF(A-EKF) (\cite{BAVDEKAR2013184}). 
\newline
Occlusion of landmarks is unavoidable in vision based methods. Several features may not be visible for a given pose due to occlusion, or due to a limited field of view. If the initial position of the robot is far apart from the target position, it may happen that the neighbourhood of these two positions do not share any common visual feature. Therefore, the robot must know if it can initiate the vision based localization with respect to the given known location or not. 
We present a multi-rate feedback based estimation approach based on the filters above to mitigate these issues.
(\cite{MultiRate_Vision_Armesto})
The highlights of the paper can be summarized as\begin{itemize}\item A state space formulation based approach is presented to suit the multi-rate homing problem in presence of uncertainties \item Observability analysis is carried out to arrive at minimum number of landmarks required for this formulation\item The efficacy of the EKF, A-EKF and PI-EKF is tested through open loop simulation studies.
\item An exercise is carried out to reach the known location(homing) using a controller that uses these estimates (\cite{ArunkumarG.K.2018}) and the closed loop performance of the scheme is evaluated.
\end{itemize}
This work builds upon the formulation presented by \cite{Misha_Med}.
The organization of this paper is as follows: The next section of this paper formulates the state space realization for the estimation exercise. The observability analysis, which gives us a sense of minimum number of landmarks to be used for the formulation, is presented in the section \ref{Section:Observability_analysis}. The estimators used in this exercise are briefly described in the section \ref{Section_Estimators_Used}. The simulation studies are presented in the section \ref{Simulation_Studies}. The section \ref{Section_Conclusion} concludes this paper.  

\section{State Space Formulation}\label{Section:Problem_Formulation}
Consider a unicycle robot moving on a 2-D plane as shown in the figure \ref{fig_model}.
Let O denotes the known target position, which, without loss of generality, is considered as the origin of the X-Y reference frame.
Let P denote the instantaneous position of the unicycle. Polar coordinates are used to denote the position of the robot and the landmarks.  
In this work, $R$ denotes the range of the robot and $\alpha$ denotes the orientation of the robot. 
Let $L_i$ be the position of the $i^{th}$ landmark as shown in the figure \ref{fig_model}. The unicycle is equipped with a monocular panoramic vision, and, in our exercise, the landmarks typically correspond to a distinct visual feature in the environment. There are $q$ landmarks in the environment.
Bearing angles of all the landmarks w.r.t the positive X-axis, at the home position $O$, are calculated and stored \textit{a-priori}, and are denoted by $\beta_1^*$, $\beta_2^*$, \dots $\beta_q^*$ respectively. The distance of the $i^{th}$ landmark from the target position $D_i^*$ is not necessarily known (particularly for the case with monocular vision), and, can be easily estimated using relations provided later in this paper. At every time instant, the robot measures the bearing angles from each landmarks as shown in the figure \ref{fig_model}. They are calculated with reference to the velocity vector $\vec{v}$ and are given by $\beta_1$, $\beta_2$, \dots $\beta_q$ respectively. The convention used here is to consider all the angles positive in an anti-clockwise direction. In our applications, the robot has to visit the landmark positions. In this paper, an augmented state space formulation is used by defining the state vector as :
\begin{figure}[bt]
\centering
\includegraphics[scale=0.5]{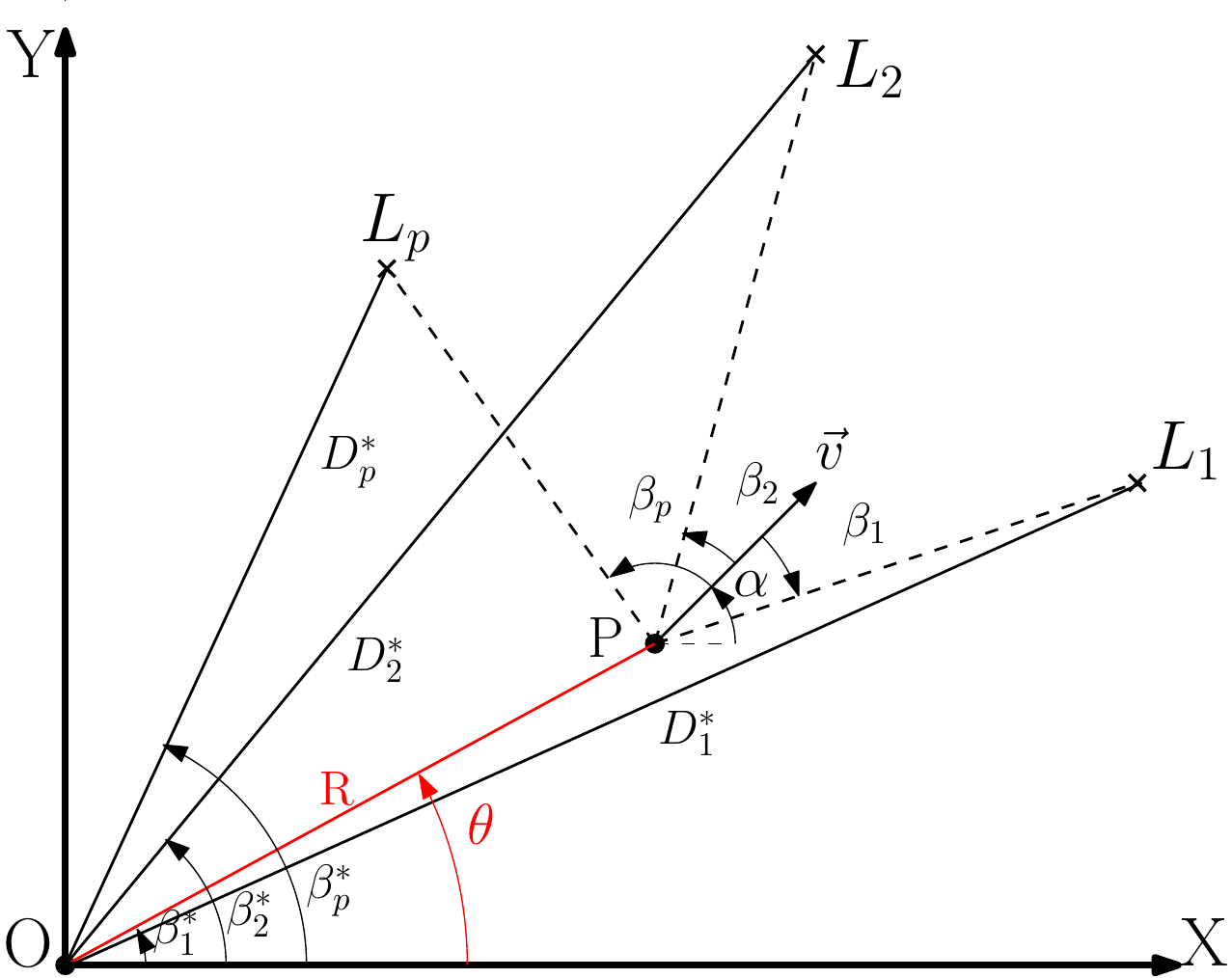}
\centering
\caption{Planar view describing the system}
\label{fig_model}
\end{figure}
\begin{equation*}
X =\begin{bmatrix}
R \ \ \ \
\theta \ \ \ \ 
\alpha \ \ \ \ 
\beta_{1} \ \ \ \ 
\beta_{2} \ \ \ \  
\ldots \ \ \ \ 
\beta_{q} 
\end{bmatrix}^{\textsf {T}}
\end{equation*}
Typically, the bearing angles of the landmarks, $\beta_i$, which are measured, are used in the control laws and such a formulation arrives at more accurate estimates of these angles.
The unicycle model, in polar co-ordinates is described by:
\begin{equation}\label{law}
    \begin{split}
    \dfrac{dR}{dt} = v \cos(\alpha - \theta) \\
    \dfrac{d\theta}{dt} = \frac{v \sin(\alpha - \theta) }{R}
    \end{split}
\end{equation}
where $v$ is linear velocity of the robot. Using the geometry in the figure \ref{fig_model} ( \cite{Misha_Med}), the following relation is established : \begin{align}
\dfrac{R}{\sin(\beta_{i}^{*}-(\beta_{i}+\alpha))} &= \dfrac{D_{i}^{*}}{\sin(\pi-(\theta-(\beta_{i}+\alpha)))}, \\
D_{i}^{*} &= R\dfrac{\sin(\theta-(\beta_{i}+\alpha))}{\sin(\beta_{i}^{*}-(\beta_{i}+\alpha))} \label{derivation}
\end{align}
for $i=1,2,3 \ldots q$. Note that $D_i^{*}$ for $i=1,2,3 \ldots q$ are constants,and are not measured. Typically, $\beta_i$, $i=1,2,3 \ldots q$  are calculated as a part of an offline exercise and these relations help in estimating $D_i^{*}$. 
The rate of change of the bearing angle with respect to time $t$ can be expressed using (\ref{law}) and (\ref{derivation}) as:
\begin{equation}
\dot{\beta_{i}} = \dfrac{-v\sin(\beta_{i})}{R\cos(\theta - (\beta_{i} + \alpha))-D_{i}^{*}\cos(\beta_{i}^{*}-(\beta_{i}+\alpha))} -\dot{\alpha} \label{diff}
\end{equation}
where $D_{i}^{*}$ is given by (\ref{derivation}). The systems of equations (\ref{law}) and (\ref{diff}) can be used to arrive at a state space representation of the form :
\begin{equation*}
 \begin{split}
     \dfrac{dx}{dt} = g(X,U)\\ \smallskip 
     y = h(x) 
     \label{sys_eq}
 \end{split}
 \end{equation*}
  where $U = [v$ $\omega]^{\textsf {T}}$, $v$ is the linear velocity of the robot and $\omega$ is its angular velocity. 
 \begin{align} 
g(X,U)&=\left(
\begin{array}{cc}
\cos(\alpha-\theta)&0\\
\dfrac{\sin(\alpha-\theta)}{R}&0\\
0&1\\
\dfrac{-\sin(\beta_{1})}{RC_1-D_{1}^{*}C_1^{*}}&-1\\
\dfrac{-\sin(\beta_{2})}{RC_2-D_{2}^{*}C_2^{*}}&-1\\
\vdots & \vdots \\ \medskip
\dfrac{-\sin(\beta_{q})}{RC_q-D_{q}^{*}C_q^{*}}&-1\\
\end{array}
\right) U \label{dotX}
\\ &= \ \ \ \ \ \ \ \ \ \ [g_1 \ \ \ \ \ \ \ \ \ g_2]\ \ U \dots say 
\end{align}
where, for brevity, $\cos(\theta-(\beta_i+\alpha))$, $ \cos(\beta_{i}^{*}-(\beta_{i}+\alpha))$, $\cos(\beta_i)$, $\sin(\theta-(\beta_i+\alpha))$, $\sin(\beta_{i}^{*}-(\beta_{i}+\alpha))$, $\sin(\beta_i)$ are concisely denoted as $C_i$, $C_i^{*}$, $c_i$, $S_i$, $S_i^{*}$, $s_i$, respectively.
The measurement equation is given by:
\begin{equation}\label{Y}
     Y =h(x)
     \begin{bmatrix}
         \beta_{1} \\ 
         \beta_{2} \\ 
         \vdots \\
        \beta_{q} 
     \end{bmatrix} = HX
     \end{equation}
     \begin{equation}
     H = [\Bar{0}_{q\times 3} \ \ I_q] 
    \end{equation}
    where $\Bar{0}_{q\times 3} \in \mathbb R^{3\times q}$ is the null matrix and $I_q\in \mathbb R^{q\times q}$ is the identity matrix. 
In this exercise, we consider that the robot is moving with a constant linear velocity. The robot is controlled by changing its angular velocity $\omega$. 
\section{Observability Analysis}\label{Section:Observability_analysis}
Our state space realization (\ref{dotX})-(\ref{Y}) consists of $3+q$ states but only the last $q$ of these states: $ \beta_{1}, \beta_{2}, \dots \beta_{q} $ are obtained from the images. We are interested in estimating the robot's configuration along with estimating the angles, $\beta_{i}, i \in\{1,2,\ldots, q\}$ using these measurements. We are also interested in arriving at the value of the minimum number of landmarks required in this formulation such that it is fully observable. 
Our state space realization (\ref{dotX},\ref{Y}) is of the form:
\begin{equation}
\dot X = g(X,U)  \qquad \qquad Y = h(X)= HX
\end{equation}
We are interested in investigating the local observability of the system. We use the Lie derivative approach for this investigation (\cite{Hermann_Nonlinear_Observability}, \cite{Misha_Med}).
 For notational simplicity, let 
$\scriptstyle(RC_i-D_{i}^{*}C_i^{*})$ ,$\frac{s_{i}(C_i-\frac{C_{i}^{*}D_{i}^{*}}{R})}{d_{i}^2}$, $\frac{Rs_{i}(S_{i}+\frac{C_{i}C_{i}^{*}}{S_{i}^{*}})}{d_{i}^2}$, $\frac{-s_{i}C_{i}^{*}}{S_{i}^{*}d_i}$, $\frac{-c_i}{d_i}-\frac{s_{i}C_{i}^{*}}{S_{i}^{*}d_i}$ be denoted as $d_i$, $J_i$, $K_i$, $P_i$, $Q_{i}$ respectively for $i \in \{1, 2, 3, \ldots q\}$.
%
The $zero^{th}$ order lie derivative of $h$ along $g$ is 
\begin{equation} \label{l1}
L^{0}h = h(X) = \left(
\begin{array}{cccc}
\beta_{1}& \beta_{2} & \ldots  & \beta_{q}
\end{array}
\right)^{\textsf {T}}
\end{equation}
\begin{equation}
\nabla L^{0}h = \nabla h(X) =  [\Bar{0}_{q\times 3} \ \ I_q] 
\end{equation}
The first order lie derivative of h along $g_1$ is 
\begin{equation} \label{l2}
\nabla L_{g_{1}}^{1}h=\left(
\begin{array}{lllllll}
J_1&K_1&P_1&Q_1&                                                                                                                                                                                                                                           0&                                                                                                                                                                                                                                           \ldots&0 \\
J_2&K_2 &P_2&0&Q_2&\ldots&0\\ 
 &\vdots & \vdots&\vdots &\vdots &\ddots & \vdots \\
J_q&K_q &P_q&0 &0&\ldots&Q_q
\end{array}
\right)
\end{equation}
\begin{equation}
L_{g_{2}}^{1} h = \frac{\partial h(X)}{\partial X}.g_{2}\\
=\left(
\begin{array}{cccc}
-1&-1&\ldots&-1
\end{array}
\right)^{\textsf {T}}
\end{equation}
\begin{equation}
\nabla L_{g_{2}}^{1}h = \Bar{0}_{q\times (q+3)}
\end{equation}
Now, consider the first three rows of the observability matrix $\mathcal{O}$ (\cite{Hermann_Nonlinear_Observability}):  
\begin{equation}
\mathcal{O}=\left(
\begin{array}{c} \label{O}
\nabla L^{0}h\\
\nabla L^{1}_{g_{1}}h\\
\nabla L^{1}_{g_{2}}h
\\ \vdots
\end{array}
\right)
\end{equation}
The observability matrix $\mathcal{O}$ in (\ref{O}) has rank $n=q+3$,(equal to our number of states), provided $q\geq 2$. Thus, our state space realization is locally observable provided that there are at-least two landmarks available.
\section{Estimators}\label{Section_Estimators_Used} We now present the formulations of the estimators used in this work.
\subsection{Multi-Rate Extended Kalman Filter Formulation} \label{Subsection:Multirate_EKF}
The conventional EKF as an observer is modified to incorporate the measurements not being available at all instants of time.
At every instant only $p \geq 2$  of the $q$ landmarks are visible to the robot. Hence, the state space formulation is reformulated as:
\begin{equation}\label{Y_p}
     Y_p =h_p(x)
     \begin{bmatrix}
         \beta_{1} \\ 
         \beta_{2} \\ 
         \vdots \\
        \beta_{p} 
     \end{bmatrix} =  H(k) X
     \end{equation}
Since the number of landmarks keep on changing at each iteration, therefore dimension of H matrix and R matrix in EKF change. Prediction and the updates are carried out as: \\
\textbf{Prediction}:-
\begin{align*}
\hat X(k\mid k-1) &= F(\hat X(k-1\mid k-1), U(k-1))\\ 
P(k\mid k-1) &= G(k)P(k-1\mid k-1)G(k)^{\textsf {T}}+ \Gamma_u Q \Gamma_u ^{\textsf {T}}
\end{align*} 
\textbf{Update} :-
\begin{align*} 
 L_p(k)= & P(k|k-1)H_p(k)^{\textsf {T}}[H_p(k)P(k| k-1)H_p(k)^{\textsf {T}} \\ & + R_p]^{-1} \\ 
e_p(k) = & [Y_p(k) - H_p(k)\hat X(k\mid k-1)] \\ 
\hat X(k\mid k) = & \hat X(k\mid k-1) + L_p(k)e_p(k) \\
P(k\mid k) = & [I - L_p(k)H_p(k)]P(k\mid k-1)
\end{align*}
where \\ { \begin{center}
  ${G}(k){=}\left[ \frac{\partial f}{\partial {X}}%
       \right]\bigg\vert_{ ({\hat{X}}(k-1|k-1),{U}(k))}$ \\ ${\Gamma_u}(k){=}\left[ \frac{\partial f}{\partial {U}}%
       \right]\bigg\vert_{ ({\hat{X}}(k-1|k-1),{U}(k))}$ 
         \end{center}}
If H is measurement matrix at initial position and home position then, $H_p(k)$ is formed with those rows of H which are
measured at the instant $k$. $R_p$
is the covariance matrix of the measured noises, corresponding to these $p$ measurements. Note that
the size of matrices and vectors with subindex $p$ varies
depending on the measured features in the output vector.
In the case where landmarks are visible, there are no measurements that are obtained for that time instant. Matrices denoted with the subindex $p$ are void and therefore no correction is performed, the state is simply predicted, providing an inter-sampling behavior (\cite{MultiRate_Vision_Armesto}). 
\\ The augmented EKF (A-EKF) formulation is a simple extension of this filter the state matrix is augmented as 
\begin{equation}
    X = [X \ \ \theta_p]^{\textsf {T}}
\end{equation}
The covariance matrix $P$ is also block diagonally appended. The variation of the parameter vector, $\theta_p$ is assumed to be governed by the random walk model (\cite{RANGEGOWDA2018411}).
\subsection{Proportional Integral-EKF}
The proportional-integral Kalman filter (PI-KF)(\cite{BODIZS2011379}, \cite{BAVDEKAR2013184}) tries to incorporate the model-plant mismatch or the errors due to linearization to carry out predictions. The original formulation is presented for linear systems and is extended in this exercise for the case wherein the state-transition map is nonlinear.  The prediction step for this filter is of the form: 
\begin{align*}
\hat X(k\mid k-1) &= F(\hat X(k-1\mid k-1), U(k-1)) + \mathcal{P}{\kappa_{i-1}}\\
\end{align*}
where $\mathcal{P} \in \mathbb{R}^{n \times n}$ is equivalent to the gain of a PI controller. ${\kappa_{i-1}} \in \mathbb{R}^n$ is the mismatch term obtained as 
\begin{equation}
    {\kappa_{i}} = {\kappa_{i-1}} + \mathcal{M}[Y(k-1)- H\hat X(k-1|k-2)]
\end{equation}
where $\mathcal{M}$ can be seen as a ``forgetting factor"(\cite{BAVDEKAR2013184}). It incorporates the impact of the past measurements on the accumulated errors. This extension of the (PI-KF) to non-linear systems is termed as (PI-EKF). The rest of the algorithm is identical to that of the EKF. The multi-rate formulation of this estimator is also along the lines of the filter described in the subsection \ref{Subsection:Multirate_EKF}.
\section{Simulation Studies}\label{Simulation_Studies}
\begin{figure}[!tb]
    \centering
    \includegraphics[width=0.99\linewidth]{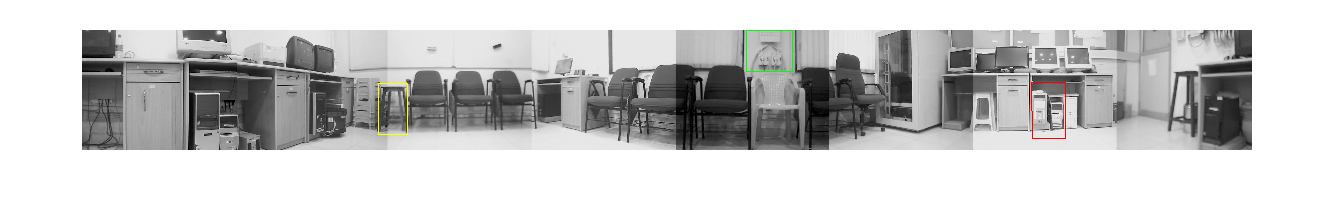}
    \caption{Representational image from the dataset (\cite{Misha_Med})} 
    \label{Dataset}
 \end{figure}

In the simulation studies, homing was used as the navigation problem which uses the estimates generated by the localization exercise. 
The simulation studies were carried out for the following cases:
\begin{enumerate}
    \item Homing using exactly known landmark information
    \item Homing using inaccurate landmark information (happens when landmark is semi-static in nature), in a multi-rate scenario.
    \item Simultaneous homing and landmark parameter estimation
    \item Open loop analysis with constant linear and angular velocity commanded: to investigate the asymptotic properties of the estimators. 
\end{enumerate}
The simulation studies are carried out in two cases: 1. Using ideal conditions where the measurements are generated using the equations (\ref{dotX},\ref{Y}) using an ode solver, and 2. Using an in-house image data-set used to generate measurements for the filter. A sample panoramic image with extracted features is shown in Figure \ref{Dataset} 

\begin{figure}[!tb]
    \centering
    \includegraphics[width=0.99\linewidth]{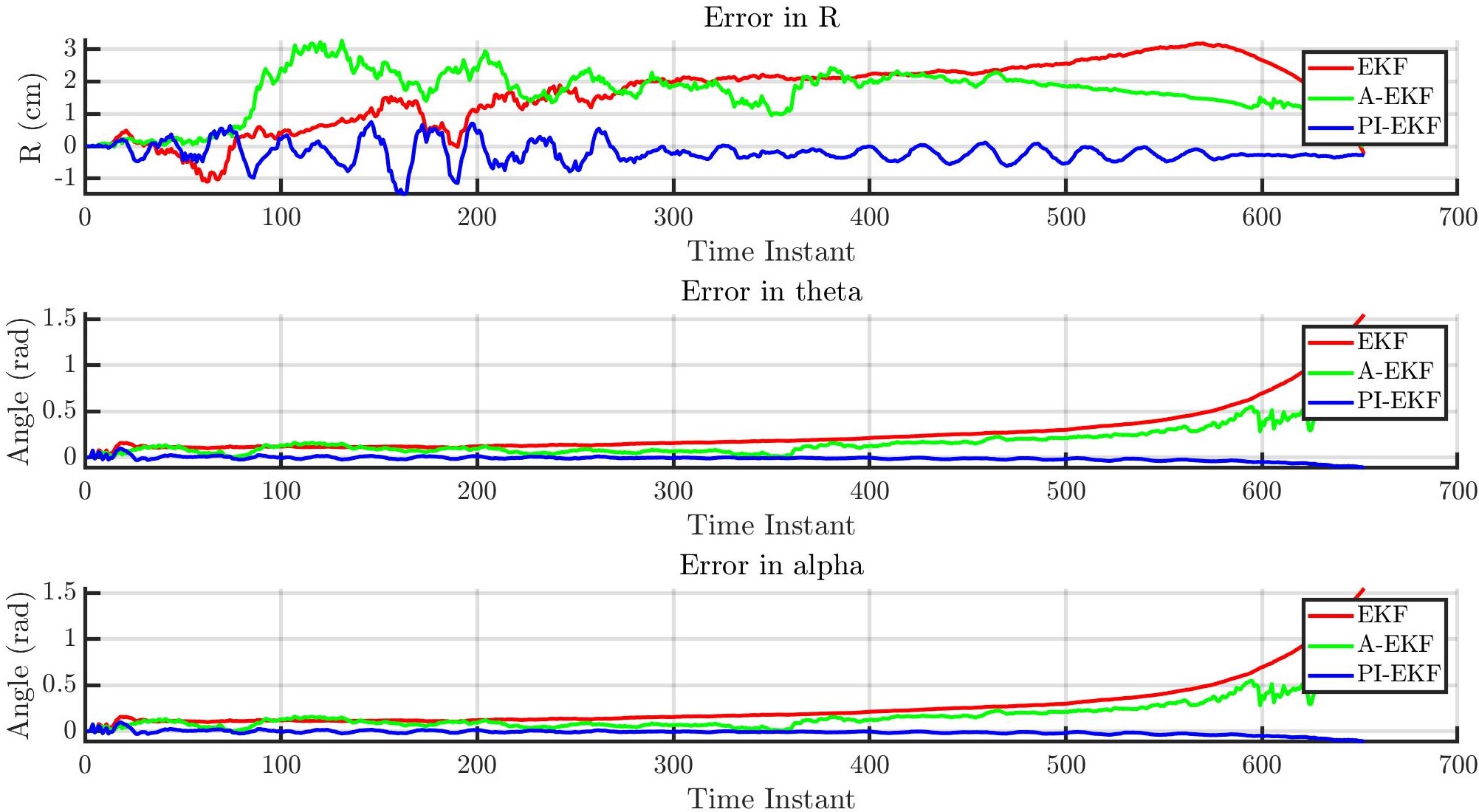}
    \caption{Estimation error (Closed Loop)} 
    \label{ClosedLoop Pose Error}
 \end{figure}

A proportional control law is used to steer the robot to the home position. The control law is
\begin{equation}
    \omega = \pi - (\hat{\alpha}(k)- \hat{\theta}(k)) 
\end{equation}
where $\hat{\theta(k)}$ and $\hat{\alpha}(k)$ are obtained from the estimators. 
\medskip \newline
\textbf{Case 2} Homing in the case where the position of the landmarks is not accurately known: In all the homing exercises, the robot commanded to move with a forward velocity of 12.5cm/sec. The robot is controlled using the angular velocity alone. The table (\ref{Table:Simulation Conditions}) gives the simulation conditions used in this exercise. The configuration of the robot is denoted by $X_r(k)= [R(k)\ \ \theta (k) \ \ \alpha(k)]^{\textsf {T}}$ 
\begin{table}[!htb]
	\centering
	\caption{Simulation Parameters}
	\label{Table:Simulation Conditions}
	\begin{tabular}{||c| c||}
	\hline 
		Parameter & Value   \\ 
		\hline
		Sampling Time & 0.05 sec \\
		$X_r(0)$   & $[75cm \ \  -0.872rad \ \ 0.349 rad]^{\textsf {T}}$\\
		$P(0|0)$ &  $\mathbf{I_n}$ \\
		$R$ & $diag(0.05)_{p \times p}$ \\
		$Q$ & $diag(0.01)_{n \times n}$\\
		$Q_d$ & $diag(0.0025)n_{ip} \times n_{ip}$ \\ 
		$\mathcal{P}$  &  0.05 $\mathbf{I_n}$\\
		$\mathcal{M}$  &  $\Bar{1}_{n \times p}$  \\
		q & 7 \\
		\hline 
	\end{tabular}
\end{table}
. 
\medskip 
The measure of performance used for comparison is: Root Mean Squared Error(RMSE).
Let the state estimation error be defined as\begin{equation}
\varepsilon(k|k)= {X}(k)- \hat{{X}}(k|k)
\end{equation}
Root Mean Squared Error$(RMSE)_i$ is  calculated for each state variable, where $(RMSE)_i$ is calculated as:
\begin{equation}
(RMSE)_i = \sqrt{\frac{1}{N} \sum_{k=1}^{N}[{x_i}(k)- \hat{{x}_i}(k|k)]^{2}}
\end{equation}
The RMSE values for the three filters have been tabulated below, while the evolution of pose error with time is depicted in Figure \ref{ClosedLoop Pose Error}. The PI-EKF results in a superior performance than the other two, however, a trade-off needs to be achieved in tuning of its parameters for guaranteeing satisfactory transient as well as steady state performance.
\begin{table}[!htb]
	\centering
	\caption{Closed Loop Performance}
	\label{Table:CLosed Loop Simulation Performance}
	\begin{tabular}{||c| c| c| c||}
	\hline 
	\diagbox[innerwidth = 3cm, height = 4ex]{Filters}{States} & $\hat{R}(cm)$ & $\hat{\theta}(rad)$ & $\hat{\alpha}(rad)$   \\ 
		\hline
		EKF & 1.9091 & 0.3840 & 0.0589 \\
		A-EKF & 1.8076 & 0.2324 & 0.0616 \\
		PI-EKF & 0.3778 & 0.0285 & 0.0271 \\
		\hline 
	\end{tabular}
\end{table}
Note that \textbf{Case 1} is a trivial case of homing with inaccurate home information available to the robot. Hence it is encompassed by the above results. The estimator is seen to converge to true state when precise information is provided, however those results are not included for the sake of brevity.
 
\begin{figure}[!tb]
    \centering
    \includegraphics[width=0.99\linewidth]{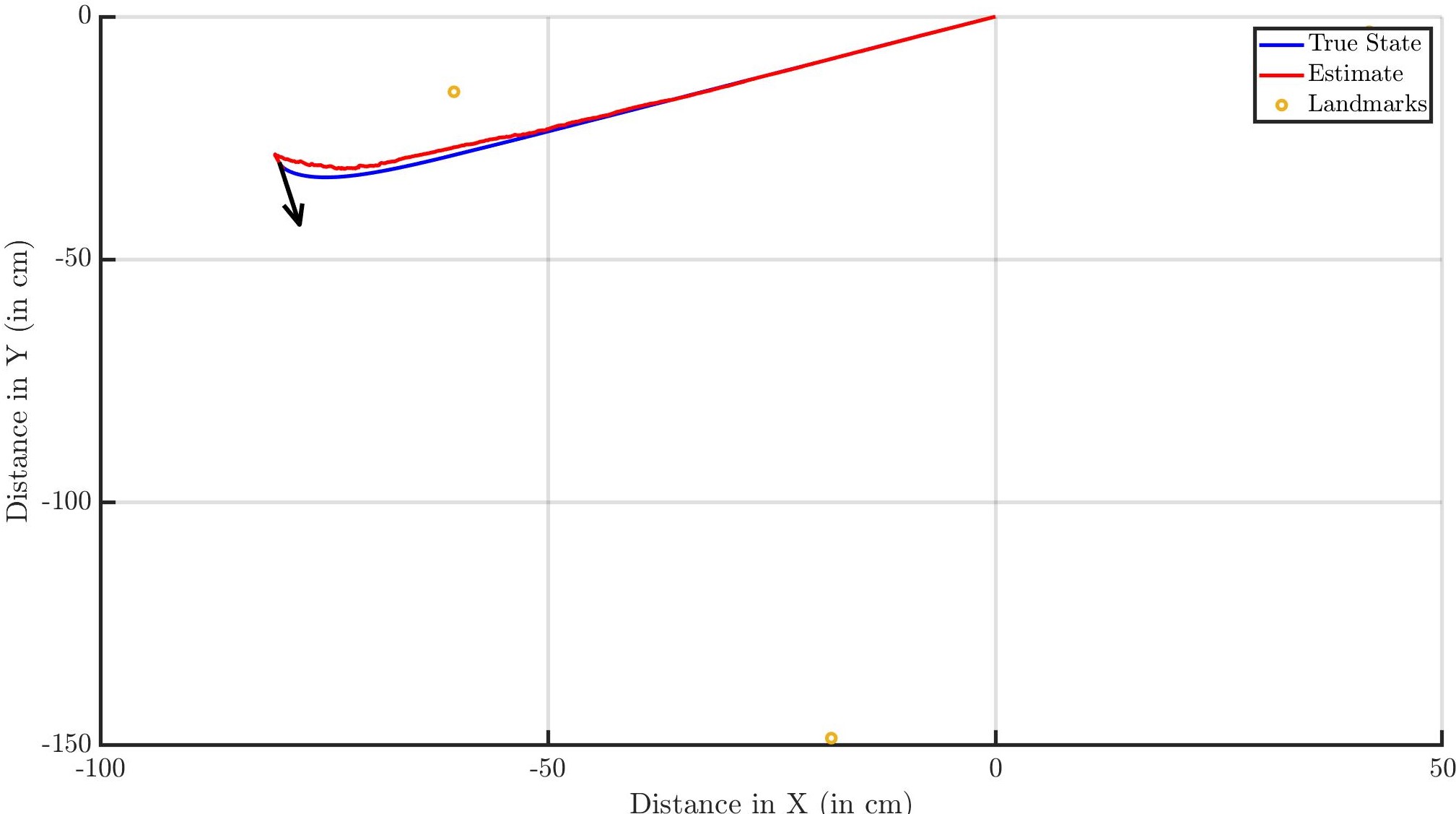}
    \caption{2-D Homing for A-EKF in presence of 1 fault}
    \label{AEKF 2D Homing}
 \end{figure}

\begin{figure}[!tb]
    \centering
    \includegraphics[width=0.99\linewidth]{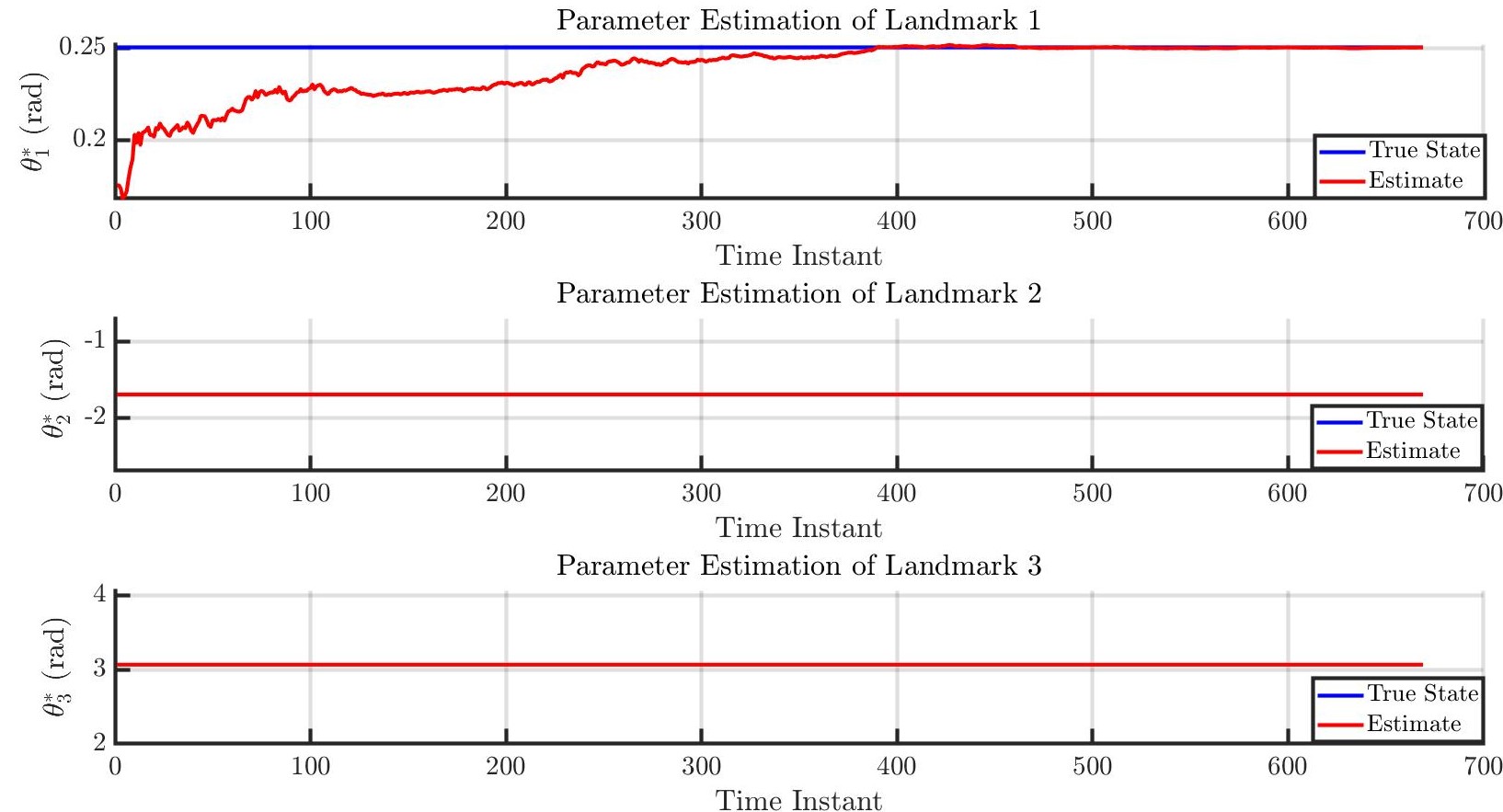}
    \caption{Estimation for a Single Parameter using A-EKF}
    \label{AEKF Parameter Estimation}
 \end{figure}

\medskip
\textbf{Case 3}: The A-EKF enables parameter estimation in conjunction with recursively estimating robot pose and bearing angle states at each time instant. This allows for the possibility to correct inaccurate landmark estimates during the navigation itself. However, performance pertaining to parameter estimation is highly sensitive to tuning. For estimation of just one parameter $\beta_1^*$, tuning is a relatively easy task. The state as well as parameter convergence can be shown in Figure \ref{AEKF 2D Homing} and in Figure \ref{AEKF Parameter Estimation}, in the case of $30\%$ error in parameter value.The initial position is marked with an arrow as shown in the figure (\ref{AEKF 2D Homing}) .
 
   As more landmark parameters are estimated (regardless of whether the initial estimate is accurate or not), the problem becomes cumbersome and sensitive to the choice of tuning and leads to inconsistent performance, both for state and parameter estimation. 
 
 \medskip
 
 \textbf{Case 4}: Open loop analysis of the estimators is crucial despite homing being guaranteed under uncertain conditions, as the homing problem is a finite time problem and estimator convergence is itself a sufficient condition for homing. A comparison is done between the three filters for robot performing circular motion, with landmark information merely being used for state prediction as part of the filter. To observe the asymptotic properties of each estimator, no faults in the knowledge of $\beta_i^*$, $i \in \{1,2, \dots p\}$ is assumed. Table \ref{Table:Open Loop Simulation Performance} shows the RMSE values for an open-loop trajectory of the robot, in the absence of any faults. 
 
 \begin{table}[!htb]
	\centering
	\caption{Closed Loop Performance (in absence of faults)}
	\label{Table:Open Loop Simulation Performance}
	\begin{tabular}{||c| c| c| c||}
	\hline 
	\diagbox[innerwidth = 3cm, height = 4ex]{Filters}{States} & $\hat{R}(cm)$ & $\hat{\theta}(rad)$ & $\hat{\alpha}(rad)$   \\ 
		\hline
		EKF & 0.4172 & 0.0174 & 0.0431 \\
		A-EKF & 0.5624 & 0.0457 & 0.0509 \\
		PI-EKF & 0.2739 & 0.0307 & 0.0386 \\
		\hline 
	\end{tabular}
\end{table}
 
\begin{figure}[!tb]
    \centering
    \includegraphics[width=0.99\linewidth]{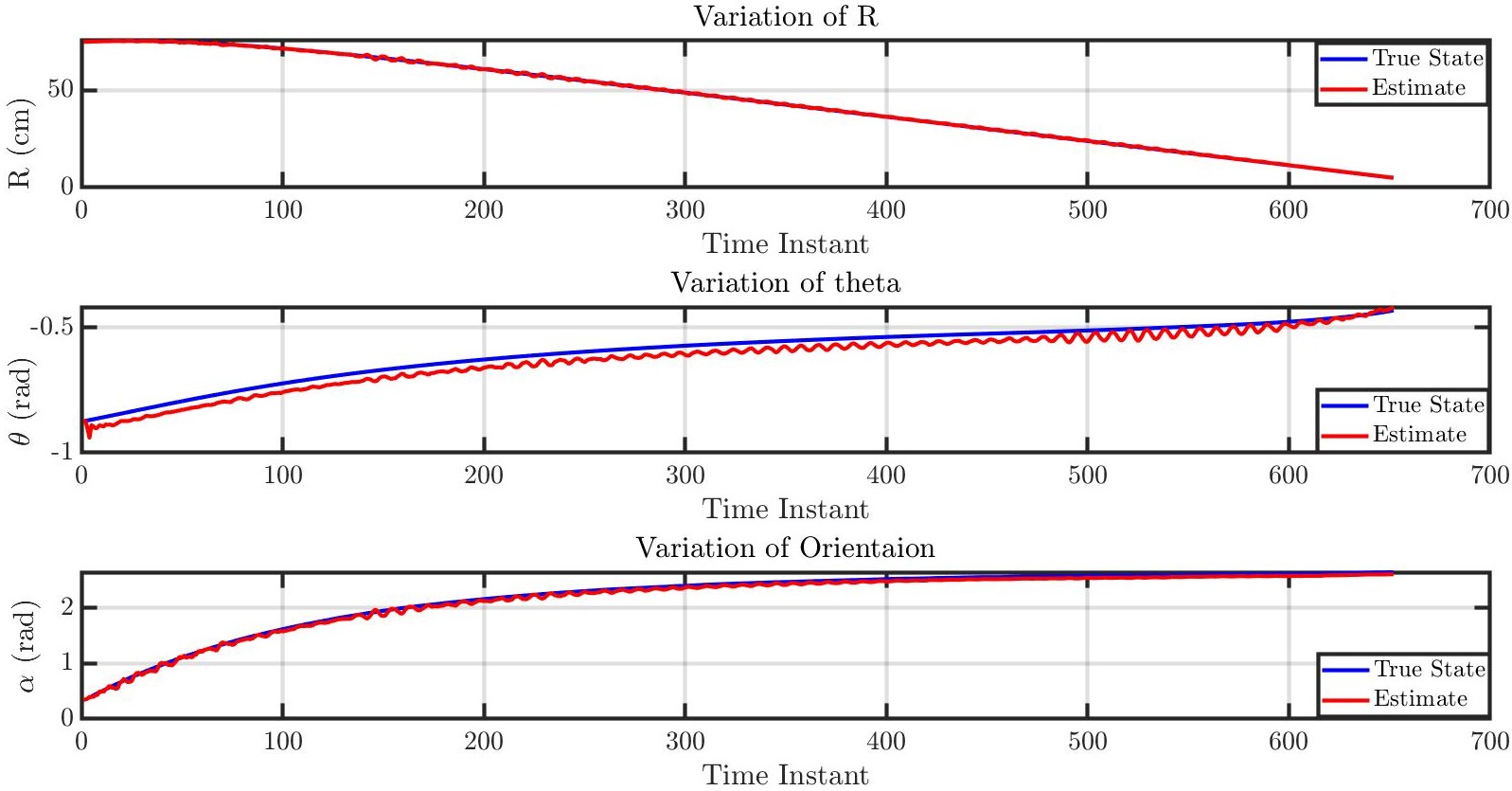}
    \caption{Performance of PI-EKF in absence of faults}
    \label{PI-EKF Pose Plots}
 \end{figure}

 Despite the significantly better performance of the PI-EKF in terms of estimation error, the interplay between the 'forgetting factor' and proportional gain for the parameter can lead to oscillatory behaviour in the estimates, as is visible in Figure \ref{PI-EKF Pose Plots}. Tuning can help achieve a balance between minimal bias and subdued oscillatory behaviour. 

 \section{Conclusions}\label{Section_Conclusion}
In this work, a state space formulation is presented to suit the multi-rate estimation problem under uncertain conditions for a homing application. Observability analysis  for the realization was carried out which shows that the choice of measurements are sufficient to estimate the configuration of the robot in the 2-d plane as well as to estimate the bearing angles of the landmarks with respect to the robot, provided that at-least two landmarks are visible to the robot. This choice of measurement also eliminates the need to measure the position of the robot with respect to the home position. 
 Simulation studies were carried out under 4 cases. Performance of three filters were evaluated: EKF, A-EKF and the PI- EKF. The conventional EKF failed perform satisfactorily for the case where there was an uncertainty in the parameter values. PI-EKF tends to show an aggressive behaviour around the steady state, and the estimation error eventually converges asymptotically even in the presence of large perturbations in the parameters. Though PI-EKF is a simple formulation, the interplay between the forgetting factor and the proportional grain plays a key role in deciding the transient performance of the filter. Tuning of the augmented EKF(A-EKF) becomes a challenging task in the multi-rate case. Although this approach appears to be simple theoretically, actual implementation of the filter strongly depends  on the manner in which the perturbation in the parameter affects the states. For the A-EKF formulation, the information transferred from the observed variables to the parameters is governed by the cross-covariances and therefore accurate information about these is crucial for a satisfactory joint performance.  In reality, the true error statics for the parameters would not be known and we have to best approximate using a simple structure like the random walk model, which poses challenges for this approach. The extension of this work could involve experimenting the mobile robots navigating through series of way points. 
\begin{ack}
The authors acknowledge Ms. Misha Gupta for her support
\end{ack}

\bibliography{ifacconf}             

\begin{thebibliography}{14}
\providecommand{\natexlab}[1]{#1}
\providecommand{\url}[1]{\texttt{#1}}
\providecommand{\urlprefix}{URL }
\expandafter\ifx\csname urlstyle\endcsname\relax
  \providecommand{\doi}[1]{doi:\discretionary{}{}{}#1}\else
  \providecommand{\doi}{doi:\discretionary{}{}{}\begingroup
  \urlstyle{rm}\Url}\fi

\bibitem[{{Armesto} et~al.(2004){Armesto}, {Chroust}, {Vincze}, and
  {Tornero}}]{MultiRate_Vision_Armesto}
{Armesto}, L., {Chroust}, S., {Vincze}, M., and {Tornero}, J. (2004).
\newblock Multi-rate fusion with vision and inertial sensors.
\newblock In \emph{IEEE International Conference on Robotics and Automation,
  2004. Proceedings. ICRA '04. 2004}, volume~1, 193--199 Vol.1.
\newblock \doi{10.1109/ROBOT.2004.1307150}.

\bibitem[{Arunkumar et~al.(2018)Arunkumar, Sabnis, and
  Vachhani}]{ArunkumarG.K.2018}
Arunkumar, G.K., Sabnis, A., and Vachhani, L. (2018).
\newblock Robust steering control for autonomous homing and its application in
  visual homing under practical conditions.
\newblock \emph{Journal of Intelligent {\&} Robotic Systems}, 89(3), 403--419.
\newblock \doi{10.1007/s10846-017-0561-2}.
\newblock \urlprefix\url{https://doi.org/10.1007/s10846-017-0561-2}.

\bibitem[{Bavdekar et~al.(2013)Bavdekar, Gopaluni, and Shah}]{BAVDEKAR2013184}
Bavdekar, V.A., Gopaluni, R.B., and Shah, S.L. (2013).
\newblock Evaluation of adaptive extended kalman filter algorithms for state
  estimation in presence of model-plant mismatch.
\newblock \emph{IFAC Proceedings Volumes}, 46(32), 184 -- 189.
\newblock \doi{https://doi.org/10.3182/20131218-3-IN-2045.00175}.
\newblock
  \urlprefix\url{http://www.sciencedirect.com/science/article/pii/S1474667015382550}.
\newblock 10th IFAC International Symposium on Dynamics and Control of Process
  Systems.

\bibitem[{Bekris et~al.(2006)Bekris, Argyros, and
  Kavraki}]{bekris2006exploiting}
Bekris, K.E., Argyros, A.A., and Kavraki, L.E. (2006).
\newblock Exploiting panoramic vision for bearing-only robot homing.
\newblock In \emph{Imaging beyond the pinhole camera}, 229--251. Springer.

\bibitem[{Bodizs et~al.(2011)Bodizs, Srinivasan, and Bonvin}]{BODIZS2011379}
Bodizs, L., Srinivasan, B., and Bonvin, D. (2011).
\newblock On the design of integral observers for unbiased output estimation in
  the presence of uncertainty.
\newblock \emph{Journal of Process Control}, 21(3), 379 -- 390.
\newblock \doi{https://doi.org/10.1016/j.jprocont.2010.11.015}.
\newblock
  \urlprefix\url{http://www.sciencedirect.com/science/article/pii/S0959152410002337}.
\newblock Thomas McAvoy Festschrift.

\bibitem[{Corke(2003)}]{corke2003mobile}
Corke, P. (2003).
\newblock Mobile robot navigation as a planar visual servoing problem.
\newblock In \emph{Robotics Research}, 361--372. Springer.

\bibitem[{{Gupta} et~al.(2017){Gupta}, {Arunkumar}, and {Vachhani}}]{Misha_Med}
{Gupta}, M., {Arunkumar}, G.K., and {Vachhani}, L. (2017).
\newblock Bearing only visual homing: Observer based approach.
\newblock In \emph{2017 25th Mediterranean Conference on Control and Automation
  (MED)}, 358--363.
\newblock \doi{10.1109/MED.2017.7984144}.

\bibitem[{{Hermann} and {Krener}(1977)}]{Hermann_Nonlinear_Observability}
{Hermann}, R. and {Krener}, A. (1977).
\newblock Nonlinear controllability and observability.
\newblock \emph{IEEE Transactions on Automatic Control}, 22(5), 728--740.
\newblock \doi{10.1109/TAC.1977.1101601}.

\bibitem[{Liu et~al.(2010)Liu, Pradalier, Chen, and Siegwart}]{liu2010bearing}
Liu, M., Pradalier, C., Chen, Q., and Siegwart, R. (2010).
\newblock A bearing-only {2D/3D}-homing method under a visual servoing
  framework.
\newblock In \emph{Robotics and Automation (ICRA), 2010 IEEE International
  Conference on}, 4062--4067. IEEE.

\bibitem[{{Meyer-Delius} et~al.(2010){Meyer-Delius}, {Hess}, {Grisetti}, and
  {Burgard}}]{Delius_Semi_Static_Environment2}
{Meyer-Delius}, D., {Hess}, J., {Grisetti}, G., and {Burgard}, W. (2010).
\newblock Temporary maps for robust localization in semi-static environments.
\newblock In \emph{2010 IEEE/RSJ International Conference on Intelligent Robots
  and Systems}, 5750--5755.
\newblock \doi{10.1109/IROS.2010.5648920}.

\bibitem[{Patwardhan et~al.(2012)Patwardhan, Narasimhan, Jagadeesan, Gopaluni,
  and Shah}]{PATWARDHAN2012933}
Patwardhan, S.C., Narasimhan, S., Jagadeesan, P., Gopaluni, B., and Shah, S.L.
  (2012).
\newblock Nonlinear bayesian state estimation: A review of recent developments.
\newblock \emph{Control Engineering Practice}, 20(10), 933 -- 953.
\newblock \doi{https://doi.org/10.1016/j.conengprac.2012.04.003}.
\newblock
  \urlprefix\url{http://www.sciencedirect.com/science/article/pii/S0967066112000871}.
\newblock 4th Symposium on Advanced Control of Industrial Processes (ADCONIP).

\bibitem[{Rangegowda et~al.(2018)Rangegowda, Valluru, Patwardhan, and
  Mukhopadhyay}]{RANGEGOWDA2018411}
Rangegowda, P.H., Valluru, J., Patwardhan, S.C., and Mukhopadhyay, S. (2018).
\newblock Simultaneous state and parameter estimation using receding-horizon
  nonlinear kalman filter.
\newblock \emph{IFAC-PapersOnLine}, 51(18), 411 -- 416.
\newblock \doi{https://doi.org/10.1016/j.ifacol.2018.09.335}.
\newblock
  \urlprefix\url{http://www.sciencedirect.com/science/article/pii/S2405896318320184}.
\newblock 10th IFAC Symposium on Advanced Control of Chemical Processes ADCHEM
  2018.

\bibitem[{{Rosen} et~al.(2016){Rosen}, {Mason}, and
  {Leonard}}]{Rosen_Semi_Static_Environmet}
{Rosen}, D.M., {Mason}, J., and {Leonard}, J.J. (2016).
\newblock Towards lifelong feature-based mapping in semi-static environments.
\newblock In \emph{2016 IEEE International Conference on Robotics and
  Automation (ICRA)}, 1063--1070.
\newblock \doi{10.1109/ICRA.2016.7487237}.

\bibitem[{Sebastian~Thrun(2005)}]{Prob_Robotics}
Sebastian~Thrun, Wolfram~Burgard, D.F. (2005).
\newblock \emph{Probabilistic Robotics}.
\newblock MIT Press, Cambridge, Massachusetts, United States.

\end{thebibliography}
                                                   







\end{document}